\documentstyle[12pt]{article}
\begin{document}
\newcommand{\etap}{\eta^\prime}
\newcommand{\vsig}{\mbox {\boldmath $\sigma$\unboldmath}}
\newcommand{\vep}{\mbox {\boldmath $\epsilon$\unboldmath}}
\bibliographystyle{unsrt}

\title{The $\eta^\prime$ Photoproduction 
Off the Nucleon In The Quark Model}
\author{Zhenping Li\thanks{e-mail: ZPLI@kelvin.phys.cmu.edu}
\\
Department of Physics, Peking University\\
Beijing, 100871, P. R. China
\\
Physics Department, Carnegie-Mellon University
\\
Pittsburgh, PA. 15213-3890
}

\maketitle

\begin{abstract}
The reaction $\gamma + p\to \etap + p$ is investigated in the
quark model approach.  The quark model predicts that the off-shell
contributions from the  resonance $S_{11}(1535)$ dominate the $\etap$
 production in the threshold region.  We find that the coulping 
constant $\alpha_{\etap NN}$ from the fit to a few total cross 
section data is small. The results of the total and differential 
cross sections are presented at different energies, which can be 
tested  in the future experiments.
If the future total cross data indeed show a resonance structure 
around 2 GeV suggested in a recent study, it would be an evidence
of a non $qqq$ structure in its wavefunction.
\end{abstract}

\vskip .5cm
PACS Numbers: 24.85.+p, 12.39.-x, 13.60.Le

\newpage

There had been a vacuum in investigating the $\etap$ photoproduction
\begin{equation}\label{10}
\gamma + p \to \etap + p
\end{equation}
since the early experiments by 
Aachen-Berlin-Bonn-Hamburg-Heidelburg -M\"unchen (ABBHHM) 
collaboration\cite{abbhhm} in the sixties and 
Aachen-Hamburg-Heidelburg -M\"unchen (AHHM) collaboration\cite{ahhm}
 in the seventies,
and no theoretical attempt was made in studying this reaction.
The construction of the continuous electron beam accellerator facility
(CEBAF) has revived the interests in this reaction.  The new experiment
proposed\cite{rechi} at CEBAF will generate more systematic data 
with much better  energy resolution and statistics.

There is little knowledge on the reaction mechanism of $\etap$
photoproductions, except several studies\cite{etap} suggesting that the
$\etap NN$ coupling constant should be small. Because the threshold 
energy of this reaction is above the second resonance region,  
data from this reaction may provide us an important probe into the structure
of the resonances around $1.9 \sim 2.0$ GeV. This was emphasized in the
recent investigation\cite{rpi} by the RPI group, in which the 
effective Lagrangian approach was used.   However, there is a large 
truncation of the model  space in the effective Lagrangian approach, 
in which the off-shell contributions from the resonances in $1.5 \sim 1.7$ 
GeV are excluded.  It remains to be seen whether the resonances around 2.0 
GeV, in particular the resonance $N^*(2080)$, would be as dominant as 
suggested in Ref. \cite{rpi}.  

In this note, we investigate  the  $\etap$ photoproduction in the quark 
model.  This is a natrual extension of the framework developed recently
\cite{zpli1,zpli2} for meson photoproductions of the nucleon.  It begins
 with the low energy QCD Lagrangian, in which the $\pi$, $\eta$ and 
$K$ are regarded as Goldstone bosons so that  their interaction with 
the quarks inside 
hadrons are invariant under the chiral transformation.  Our studies in 
Kaon and $\eta$ photoproductions have shown that the quark model gives 
a very good description of the meson photoproductions with far less free 
parameters.  Although the $\etap$ could not be regarded as a Goldstone
boson, it has the same quantum numbers as those of the $\eta$.  Thus,
the reaction mechanism for both photoproductions should be very similar,
and the transition amplitudes for both photoproductions should have 
the same expressions for a pointlike $\etap$.  Of course, it may 
be unrealistic to treat the $\etap$ as a pointlike particle.  
Nonetheless, we expect that the qualitative features from this 
investigation will survive the finite size effects.

We shall discuss briefly the quark model approach to the $\etap$ 
photoproductions of the nucleon, as the detailed formalism in the 
quark model has been given in Ref.\cite{zpli2}.  
Because the $\etap$ is a charge neutral particle, there are only two 
components in the quark model for the $\etap$ photoproduction; the s- and
u- channel contributions from the baryon resonances.  The 
transition amplitude is 
\begin{equation}\label{1}
{\cal M}_{if}={\cal M}_s+{\cal M}_{u}.
\end{equation}
The t-channel contributions, such as the $\rho$ and $\omega$ 
exchanges that played an important role in the effective Lagrangian
approach, are excluded here with the input of the duality 
hypothesis\cite{dolen,wjs}.   Whether this hypothesis 
is valid remains to be investigated for the $\etap$ photoproduction,
although the studies suggest that it may be true for the Kaon
photoproduction\cite{wjs}.

The u-channel ${\cal M}_u$ in Eq. \ref{1} includes the contributions
from the nucleon and the excited baryons with isospin 1/2,  
of which the formulae have been given in Ref. \cite{zpli2}.  The excited 
baryons are treated as degenerate so that their total contributions
can be written in a compact form in the quark model.  This
is a good  approximation since the contributions from the u-channels 
resonances are not sensitive to their precise mass positions. 
The transition amplitude ${\cal M}_s$ in Eq. 1 is\cite{zpli2}
\begin{equation}\label{2}
{\cal M}_s=\sum_R \frac {2M_R}{s-M_R(M_R-i\Gamma_R ({\bf q}))}
e^{-\frac {{\bf k}^2+{\bf q}^2}{6\alpha^2}} {\cal O}_R,
\end{equation}
where the resonance $R$ has the mass $M_R$ and total width $\Gamma_R$,
 ${\bf k}$ and ${\bf q}$ are the momenta of  incoming photons
and outgoing mesons, and $\sqrt {s}$ is the total energy of the
system.  The operator ${\cal O}_R$ in Eq. \ref{2} depends on the 
structure of resonances.  It is devided into 
two groups; the s-channel resonances below 2 GeV and those above 2 GeV 
that could be regarded as continuum contributions.
The electromagnetic transitions of s-channel baryon resonances and
their meson decays have been investigated extensively in the quark 
model\cite{cko,LY,simon,close} in terms of the helicity and
the meson decay amplitudes.  These transition amplitudes for s-channel
resonances below 2 GeV have been translated into the standard
 CGLN\cite{cgln}  amplitudes in Refs. \cite{zpli1,zpli2} for 
the proton target and
\cite{zpli3} for the neutron target in the harmonic oscillator basis.
The advantage of the standard CGLN variables is that the kinematics
of the meson photoproductions has been thoroughly investigated\cite{tabakin}, 
the various observables of the meson photoproductions could be easily 
evaluated in terms of these amplitudes.
Those resonances above 2 GeV are treated as degenerate, since few 
experimental information is available on those resonances.
Qualitatively, we find that the resonances with higher partial waves
have the largest contributions as the energy increases.  Thus, we
write the total contributions from the resonances belonging to the same 
harmonic oscillator shell in a compact form, and the mass and total 
width of the high spin states, such as $G_{17}(2190)$ for $n=3$
harmonic oscillator shell, are used.  The contributions from the 
resonances above 2 GeV are particularly important for the $\etap$
photoproduction, since the threshold energy is higher than those
in the Kaon and $\eta$ photoproductions.  We have included the 
resonances belonging $n=4$ and $n=5$ with the resonance mass up 
to 2.6 GeV.  Thus, our calculation becomes less reliable for 
$E_{lab}\ge 3.0$ GeV, in which the continuum contributions
in our calculation are nolonger adequate.

We assume that the relative strength and phases of each term in s- and
u-channels are determined by the quark model wavefunction
with exact $SU(6)\otimes O(3)$ limit.  The masses and decay widths
of the s-channel baryon resonances are obtained from the recent particle
data group\cite{pdg94}.  
Thus, there are three parameters in this calculation; the coupling
constant $\alpha_{\etap NN}$, the constituent quark masses $m_q$ for
up or down quarks, and the parameter 
$\alpha^2$ from the harmonic oscillator wavefunctions in the quark
model.  The quark masses $m_q$ and the parameter $\alpha^2$ are
well determined in the quark model, they are
\begin{eqnarray}\label{20}
m_{u}=m_{d}=0.34 & \quad & \mbox{GeV} \nonumber \\
\alpha^2=0.16 & \quad & \mbox{GeV}^2 .
\end{eqnarray} 
The finite size effects of the $\etap$ could be partially taken into
account by adjusting the parameter $\alpha^2$, which we leave it to
the future investigation.  This leaves only {\bf one} free parameter, 
the coupling constant  $\alpha_{\etap NN}$, to be determined 
in our calculation.

In Fig. 1, we show the result of our calculation for the total 
cross sections of the $\etap$ photoproduction off the proton target.  
The coupling constant $\alpha_{\etap NN}$ is $0.35$ from the 
fit to a few total cross section data, which is indeed 
consistent with the recent studies\cite{etap} suggesting it to 
be small.  Of course, there is a large uncertainty due to the 
poor quality of the data. Our result does not exhibit the 
dominance of any particular resonance around 2 GeV region.  
This can be understood by the relative strength of the CGLN 
amplitudes ${\cal O}_R$ in Eq. \ref{2} between the resonance $S_{11}(1535)$
and the resonances around 2.0 GeV in the quark model.
The operator ${\cal O}_R$ for the resonance $S_{11}(1535)$ is
\begin{equation}\label{3}
{\cal O}_{S_{11}(1535)}=\frac {k}{3}\left ( \frac {\omega_{\etap}}{m_q}+
\frac {{\bf A}\cdot {\bf q}}
{3\alpha^2}\right )\left (1+\frac {k}{2m_q}\right ) \vsig \cdot \vep 
\end{equation}
where 
\begin{equation}\label{54}
{\bf A}=-\left (\frac {\omega_{\etap}}{E_f+M_N}+1\right ){\bf q},
\end{equation}
and $\vsig$ and $\vep$ are the total spin operator and the polarization
vector of the incoming photons.
According to the quark model classification, the $S$ or $D$ wave 
resonances around 2 GeV belong to $n=3$ in the harmonic oscillator
basis.  The operator ${\cal O}_R$ for the $n=3$ resonances is
\begin{eqnarray}\label{4}
{\cal O}_{n=3} = -\frac {1}{12m_q}i\vsig \cdot {\bf A}\vsig 
\cdot (\vep\times {\bf k})\left (\frac {{\bf k}
\cdot {\bf q}}{3\alpha^2}\right )^3 \nonumber \\
+\frac 1{6}\left [\frac {\omega_{\etap} k}{m_q}\left (1+\frac 
{k}{2m_q}\right )\vsig \cdot  \vep  + \frac
{k}{\alpha^2}\vsig\cdot {\bf A}\vep\cdot
{\bf q}\right ]\left (\frac {{\bf k}\cdot {\bf q}}
{3\alpha^2}\right )^2 \nonumber \\ +\frac {\omega_{\etap} k}{9\alpha^2 m_q}
\vsig\cdot {\bf k}\vep\cdot {\bf q} \left (\frac {{\bf k}\cdot {\bf q}}{3
\alpha^2} \right ).
\end{eqnarray}
  Because the spatial 
wavefunctions for the $S$ and $D$ wave resonances are orthogonal to that
of the $S_{11}(1535)$,  the dependence on ${\bf k}$ and ${\bf q}$ of the 
CLGN amplitudes for the $S$ and $D$ wave resonances around 2 GeV should 
be very different from those for $S_{11}(1535)$ and $D_{13}(1520)$.  
Indeed, Eq. \ref{4} shows that the amplitude for the $S$ and $D$ wave 
resonances with $n=3$ is at least proportional to ${\bf q}^2$ comparing
to the ${\bf q}$ dependence of the amplitude of the $S_{11}(1535)$
in Eq. \ref{3}.  Thus, the $S$ and $D$ wave resonances around 2 GeV give 
little contribution to the $\etap$ productions in the threhsold region.
Moreover, Eq. \ref{4} represents the 
sum of every resonance with $n=3$, and the $G$ wave resonance has the 
largest amplitude among these resonances.  The magnitude of the CGLN
amplitudes for $S$ and $D$ wave resonances is even smaller that that
for the $S_{11}(1535)$, 
since the magnitude of Eq. \ref{4} is about 10 times smaller than
that of Eq. \ref{3}.  Thus, 
the dominance from a  particular resonance around 2 GeV in the 
threshold region,  similar to the
dominance of the resonance $S_{11}(1535)$ in the threshold region
of the $\eta$ photoproduction, is not expected, unless a resonance
has a non $qqq$ structure in its wavefunction. This suggests 
that the s-wave  resonances in $1.5 \sim 1.7$ GeV region, 
in particular $S_{11}(1535)$, play quite important role in the threshold
 region of the $\etap$ photoproduction.  To show the importance of the 
off-shell contributions from the resonance $S_{11}(1535)$, we show
the result of our calculation without the contribution from the 
resonance $S_{11}(1535)$; the resulting total cross section is 
only about 20\% of that with the the contribution from the 
$S_{11}(1535)$.   The differential cross sections in Fig. 2 
also show the S-wave dominance in  the threshold region,  although it 
is slightly forward peaked.  Thus, the off-shell contributions
from the $S_{11}(1535)$ is dominant in the threshold region 
of the $\etap$ production.  This result represents an important 
prediction of the quark model,  because the relative strength and 
phases of the CGLN amplitudes for each resonance are determined 
by the quark model wavefunctions.

Fig. 1 also shows a small bump  around $E_{lab}=2.1$ GeV region, 
and it comes from the resonances $G_{17}(2190)$ and $H_{19}(2220)$.  
This suggests that if there is indeed a dominance from the resonance 
$N^*(2080)$ with the magnitude shown in Ref. \cite{rpi}, its magnitude 
of the CGLN amplitude  should be much larger than that expected in the 
quark model, and it is only possible for a non $qqq$ state.  
Thus, the search of the missing resonances requires more elaborate
studies of the polarization data proposed in Ref. \cite{frank}, 
as the quark model predicts that $S$ and $D$ wave resonances around
2 GeV do not make significant contributions to total and differential
cross sections of the $\etap$ 
photoproduction.

It should be also pointed out that the resonance structure around 
$W=2$ GeV is far from  being established in the $\etap$ photoproduction
data.  There are total four points of data in $E_{lab}\le 2.4$ GeV from 
ABBHHM collaboration (triangle) and from AHHM collaboration (square).
On the surface, the data seem to show a resonance structure
with very large uncertainties.  However, if one separates the two sets
of the data by ABBHHM and AHHM collaborations, the resonance structure
disappears. Consequently,  the total cross section decreases monotonously
as the energy increases, consistent with the results of the quark model 
calculations.  Moreover, the dominance of $D_{13}(2080)$ shown 
in Ref. \cite{rpi} requires further support from the differetial 
cross section data, which do not exist at present.  

Thus, our investigation here provides an interesting challenge to the
future experiments.  A strong peak centered around 1.9 GeV shown
in Ref. \cite{rpi} in the total cross section data suggests the 
existence of a resonance that should not be a normal $qqq$ state. 
On the other hand,   the contribution from the resonance $S_{11}(1535)$
is determined by the off-shell behaviour of its CGLN amplitudes.
Whether the CGLN amplitudes from the quark model give a good description
of the off-shell behaviour of baryon resonances remains to be 
seen,  which are usually evaluated on-shell. 
The differential and total cross section data in the threshold
region would provide an important test to the off-shell behaviour
of the CGLN amplitudes from the quark model.  

The author acknowledges the discussions with N. Mukhopaydhyay. 
This work was supported in part
by the U.S. National Science Foundation grant PHY-9023586.

\subsection*{Figure Caption}
\begin{enumerate}
\item The total cross sections for $\gamma + p\to \etap + p$.
The difference between the solid and dashed lines represents
the importance of the contribution from the resonance $S_{11}(1535)$.
 The experimental data are from Refs. \cite{abbhhm} (triangle) 
and \cite{ahhm} (square).  See text.

\item The differential cross sections at $E_{lab}=1.6$ GeV 
(solid), $1.8$ GeV (dashed) and 2.0 GeV (dot-dashed).
\end{enumerate}




\end{document}